\documentclass[preprint,showpacs,preprintnumbers,amsmath,amssymb]{revtex4}

\usepackage{graphicx}
\usepackage{dcolumn}
\usepackage{bm}

\begin{document}


\title{Bosonization of Matter $\beta\gamma$ CFT, Non-compact Coordinate and Noncritical String Theory}

\author{Bom Soo Kim}
\affiliation{Berkeley Center for Theoretical Physics, University of California, Berkeley, CA 94720\\
Theoretical Physics Group, Lawrence Berkeley National Laboratory, Berkeley, CA 94720}

\date{\today}

\begin{abstract}
We consider a (supersymmetric) bosonization of general commuting matter $\beta\gamma$ CFT. 
Unlike the conventional ghost $\beta\gamma$ CFT bosonization, which is typically described in terms of one 
scalar and one set of $\xi\eta$ CFT, the matter $\beta\gamma$ CFT is naturally bosonized to two scalar CFTs. 
Surprisingly, there exists an independent subsector, which satisfies a constraint necessary for consistent 
bosonization, while the usual sector has a description in terms of two compact coordinates. 
The subsector itself is a complete system which is equivalent to the original $\beta\gamma$ CFT. 
Furthermore, a non-compact coordinate naturally emerges in this ``non-compact subsector''. 
As an immediate application, we consider the bosonization of matter $\beta\gamma$ CFT in the context of full string theory and  
find the connection to noncritical string theories in the light-cone gauge.    

\end{abstract}

\maketitle


{\it Introduction.} --- 
Bosonizations in two dimensions are a remarkable tool for investigating quantum field theories 
\cite{Coleman:1974bu} 
and string theories \cite{Green:1987mn, Polchinski:1998rq}. For the bosonization of fermions or ghost fields 
propagating on a circle, the corresponding bosonic field has half integral eigenvalues in its momentum operator 
or ghost number operator, respectively. Thus the zero mode of the bosonized field is an angular variable. The bosonized field is  
interpreted as a compact coordinate in string theory.  \\
                \indent In this letter we would like to investigate a bosonization of the commuting matter 
$\beta\gamma$ CFT which was considered recently\cite{Kim:2007pc, Gomis:2000bd}. 
This matter CFT is fully bosonized in terms of two scalar $\phi\chi$ CFTs. We call this as 
``double bosonization''\cite{footnote1}
in contrast to ``single bosonization'' with a scalar and an anti-commuting $\xi\eta$ CFT. 
Along the way, we notice clear discrepancies in zero modes between $\beta\gamma$ CFT and $\phi\chi$ CFTs. 
The double bosonized form reveals natural solutions to these mismatches by providing a constraint which, 
in turn, removes compactness condition for the bosonized $\phi\chi$ fields. 
Remarkably, a non-compact coordinate emerges from the double bosonization. \\
\indent There exist important discrepancies in the zero modes for the ghost $\beta\gamma$ CFT. This was already noticed in the single 
bosonization language in the original paper \cite{Friedan:1985ge}. 
It was mentioned in the paper as ``{\it It is curious and essential for future developments that 
the zero mode $\xi_0$ never appears in the $b,c$ algebra of Bose systems; only derivatives of $\xi$ are needed.}'' 
There were further attempts to solve these issues with different bosonization form and/or in terms of BRST 
cohomology\cite{Horowitz:1988xf}. These issues were not fully 
understood in the previous works because (i) the main focus of $\beta\gamma$ bosonization was on the ghost system, 
and (ii) attempts were mostly concentrated on the single bosonization. 
Those reasons were necessary to understand the ghost structure and the superstring quantization, 
which could be done by the single bosonization in a 
very convenient way. Now we would like to fully understand the double bosonization 
of the {\it matter} $\beta\gamma$ CFT in terms of two scalar $\phi\chi$ CFTs, discrepancies in their zero modes structure 
and emergence of the non-compact coordinate.    \\                                                                        
\indent {\it Bosonization of zero modes} --- 
Holomorphic commuting matter $\beta\gamma$ CFT, with conformal weight of $\beta$ as $\lambda$, has the usual mode expansion 
\begin{equation}
\gamma (z) = \sum \frac{\gamma_m}{z^{m+1-\lambda}} \ , ~~ \beta (z) = \sum \frac{\beta_m}{z^{m+\lambda}} \ ,
\end{equation}
and their zero mode and higher modes commutation relations are 
\begin{equation}
[\gamma_0, ~\beta_0] = 1 \ , ~~ [\gamma_n,~ \beta_m] = \delta_{m+n} \ . 
\end{equation}
Here we can see that there is only one set of commutator for each integer mode $n$, including the zero mode commutator.  
Thus after the bosonization we just need the same set of commutators to have the equivalence.  \\                       
\indent We can proceed to think about the bosonization formula presented in \cite{Friedan:1985ge}
\begin{equation}
\gamma = e^{\phi} \eta = e^{\phi -\chi} \ , ~~ \beta = e^{-\phi} \partial \xi = e^{-\phi + \chi} \partial \chi \ .
\label{bosonizationformula}
\end{equation}
Conventionally the bosonization procedure has two steps. The first step has a set of fields, a scalar field $\phi$ and 
anti-commuting $\eta\xi$ CFT. 
Usually the bosonization stops at this step because it is convenient to analyze the ghost zero mode structure in terms 
of the zero modes of the $\eta\xi$ CFT. There is a well known issue related to this. The zero mode of the field $\xi$ does 
not enter the bosonization formula and the zero mode of the field $\eta$ is actually irrelevant. These zero modes 
contribute neither to superstring dynamics nor to the zero mode commutation relations. In BRST quantization, 
we essentially impose a condition by hand to eliminate half of the Hilbert space produced by the ghost zero mode $\eta_0$. 
Some attempts has been made to resolve the issues on zero modes with the full bosonization in terms of two 
scalar fields $\phi$ and $\chi$, where $\partial \beta$ and $\eta$ are exchanged. Then the BRST structure  
is different due to the distinguished role of ghost $\gamma$ field \cite{Horowitz:1988xf}. 
In this bosonization, ghost number can be arbitrary and every cohomology class can be written as a picture 
changed version of the fixed ghost number cohomology classes. 
Even though those efforts clarify some parts of the issues, the comment on zero modes mentioned in the introduction is left unanswered. \\
\indent Now we investigate the double bosonization formula given in (\ref{bosonizationformula}) in cylindrical coordinates, 
$(\tau, ~ 0 \leq \sigma < 2\pi)$, which is more transparent for our purpose.  
Two bosonic fields $\phi$ and $\chi$ on this coordinate with $\tau = 0$ have the mode expansions \cite{Green:1987mn} 
\begin{eqnarray}
&&\phi = \phi_0 + \sigma p_0 + i \sum_{n\neq 0} ~(1/n)~ \phi_n~ e^{-in\sigma} \ , \nonumber \\
&&\chi = \chi_0 + \sigma k_0 + i \sum_{n\neq 0} ~(1/n)~ \chi_n~ e^{-in\sigma} \ .
\label{modeexpansionincylinder}
\end{eqnarray}
Because we have two free scalars, we have two independent sets of the zero modes  
\begin{eqnarray}
&[p_0, \phi_0] = i \ , &~~~[\alpha_m^{\phi}, \alpha_n^{\phi}] = m \delta_{m+n} \ ,  \nonumber \\
&[k_0, \chi_0] = -i \ , &~~~[\alpha_m^{\chi}, \alpha_n^{\chi}] = - m \delta_{m+n} \ .
\label{zeromodealgebras}
\end{eqnarray}
Note that we have a relative sign in the commutation relations between the two scalar fields. 
The sign difference come from the fact that $\phi$ is the bosonized field of the commuting $\beta\gamma$ CFT 
and $\chi$ is the bosonized field of the anti-commuting $\xi\eta$ CFT.                                                   \\
\indent The field $\gamma$ has the double bosonization formula given in equation (\ref{bosonizationformula}) and can be written 
in terms of the modes of the bosonized fields. Concentrating on zero mode part only, we have  
\begin{eqnarray}
\gamma &=& (\cdots e^{i\phi_0} ~e^{i \sigma (p_0 +1/2)} \cdots) \nonumber 
 (\cdots e^{-i\chi_0} ~e^{-i \sigma (k_0 +1/2)} \cdots)  \\
&=& \cdots   e^{i(\phi_0 -\chi_0)} ~e^{i \sigma (p_0 +1/2 - k_0 -1/2)} \cdots  \ . 
\label{bosonizedgamma}
\end{eqnarray}
Note that $p_0$ and $k_0$ are shifted to $p_0 +1/2$ and $k_0 +1/2$, respectively,   
to ensure the change of statistics for each field $\phi$ and $\chi$. 
In the combined field, $\phi -\chi$, the factors $1/2$ cancel out in its momentum $p_0 - k_0$. 
Furthermore, we need to impose a constraint $p_0 - k_0 = n $, where $n=$ integer, because the field $\gamma$ and 
its bozonized form (\ref{bosonizedgamma}) are single valued. This restriction suggests, at least, 
the combined coordinate $\phi -\chi$ should be a compact coordinate. 
This is expected. But here comes a surprise. Momentarily, we will show that it is further required to put $n=0$, 
and then the requirement for the compact coordinate drops out for the combination $\phi_0 -\chi_0$. \\   
          \indent Let us mention a clear discrepancy of the bosonization without the requirement on $n$. 
We start off with one set of zero mode in $\beta\gamma$ CFT and end up with two sets in $\phi\chi$ CFTs after 
the double bosonization. 
It turns out that we can fix this by going to the light-cone coordinate and impose a constraint on the momenta 
of the bosonized fields, which effectively reduces the number of zero modes by half. We introduce the coordinates 
\begin{equation}
x^{\pm} = (\chi \mp \phi) / \sqrt{2}  \ ,
\end{equation}
and further impose the constraint $n=0$ as 
\begin{equation}
p^- = k_0 - p_0 = 0  \ .
\label{bosonizationcondition}
\end{equation}
Then we have the commutation relations 
\begin{equation}
[ p^+ , x^-] =   i \ , \hspace{0.3 in}  [ p^- , x^+] =  0 \ . 
\label{commutationrelationof+-}
\end{equation}
This shows that with the condition (\ref{bosonizationcondition}), one of the commutation relations is 
identically 0 and we end up just having one set of zero modes. Thus we have a correspondence between the two sides of the 
bosonization. Furthermore this frees the coordinate $\phi + \chi$ to be a non-compact coordinate and 
we can still use the zero mode $\phi_0 - \chi_0 = \tau$ in the light-cone quantization. \\       
            \indent For the $\beta\gamma$ CFT, the Hamiltonian has $L_0$ part of the energy momentum tensor together 
with the central charge, $H = L_0 - c/24$. The ordering constant in $L_0$ and the $\lambda$ dependent part of 
the central charge add up to $ -1/12 $, which is independent of $\lambda$.
\begin{equation}
H_{\beta\gamma} = \sum_{n=1}^{\infty} n ~(\beta_{-n} \gamma_n - \gamma_{-n} \beta_n ) -1/12  \ .
\end{equation} \\                                                                                                                            \indent For the bosonized fields $\phi\chi$ CFTs, the total Hamiltonian is the linear combination of the two 
Hamiltonians. $L_0$s of the fields $\phi$ and $\chi$ have a relative minus sign, while the central charges add 
up to $-2/24$  
\begin{equation}
H_{\phi\chi}
= \frac{p_0^2 - k_0^2}{2} + \sum_{n=1}^{\infty} ~(\phi_{-n} \phi_n - \chi_{-n} \chi_n ) - \frac{1}{12} \ .
\end{equation}
Here we can further check that we need to impose the condition (\ref{bosonizationcondition}) to have the 
correspondence between the two CFTs through double bosonization. Then we can check that these two Hamiltonians are 
the same because the factor $n$ in $H_{\beta\gamma}$ comes from the different normalizations of the commutators between 
the two CFTs. The equivalence of the partition functions can be checked trivially because the 
Hamiltonians, on top of the higher modes structures, are the same. \\
   \indent {\it Supersymmetric case} --- 
Superghosts are important in the context of the superstring quantization and are well understood 
through the bosonization with a triplet of scalar fields $\sigma, \hat{\phi}, \chi$\cite{Friedan:1985ge}. 
Manifest supersymmetric bosonization of the superghosts was also constructed with two bosons and one fermion 
for consistent supersymmetric string calculation with manifest superconformal 
symmetry\cite{Martinec:1988bg}. It is interesting to check how a non-compact coordinate 
emerges in the bosonization of the supersymmetric version of matter $\beta\gamma$ CFT.  \\
              \indent  $\beta\gamma$ CFT and its superpartner $bc$ CFT 
can be combined in the superfields 
${\bf \Gamma}(z, \theta) = -\gamma (z) + \theta c(z)$ and ${\bf \Sigma}(z, \theta) = b(z) + \theta \beta(z)$. 
Conformal dimensions are $h_{\Gamma} = 1-\lambda$, $h_{\Sigma} = \lambda - 1/2$ for fields with arbitrary conformal weight.
Manifest sypersymmetric bosonization of the fields ${\bf \Gamma},{\bf \Sigma}$ has the form 
\cite{Martinec:1988bg}
\begin{equation}
{\bf \Gamma} = e^{\Phi} \ , 
~~{\bf \Sigma} = e^{-\Phi} D \bar{\Phi} \ , 
\label{superbosonization}
\end{equation}
in terms of two superfields $\Phi(z, \theta) = \phi + \theta \psi$ and $\bar{\Phi}(z, \theta) = \bar{\phi} 
+ \theta \bar{\psi}$. Their nonzero two point function is $\langle \Phi(1) \bar{\Phi}(2) \rangle = \ln (\hat z_{12} )$,
where $\hat{z}_{12} = z_1 - z_2 -\theta_1 \theta_2$ and $\theta_{12} = \theta_1 - \theta_2$. It is straightforward to 
check the OPE, ${\bf \Gamma}(1)  {\bf \Sigma}(2) \sim \theta_{12}/\hat{z}_{12}$. \\
              \indent  To investigate the zero mode structure, it is more convenient to change the field $\phi$ into $\phi_1$ and $\phi_2$, 
$\phi =(\phi_1 -  \phi_2)$ and similarly for the other fields. 
The nonzero OPEs are $\phi_1(z_1) \phi_1(z_2) =  \log(z_{12}) = - \phi_2(z_1) \phi_2(z_2)$.
The mode expansions and commutators of these fields are similar to the expressions (\ref{modeexpansionincylinder}) and 
(\ref{zeromodealgebras}).
The bosonization formula (\ref{superbosonization}) requires equivalence of the fields $\Gamma$ 
\begin{equation}
{\bf \Gamma}(z, \theta) = e^{\Phi} = e^{(\phi_1 -  \phi_2)}\left(1 + \theta (\psi_1 - \psi_2) \right)\end{equation}
under the shift of $e^{2\pi}$ in the world sheet coordinate $z$, $\Gamma(ze^{2\pi}) = \Gamma(z)$. 
With the same reason given in the previous section, we are required to impose the condition that 
the momenta of the fileds $\phi_1$ and $\phi_2$, equivalent to $\phi$ and $\chi$ for the bosonic case, 
should be the same, {\it i.e.,} $p^- = p_{10} - p_{20} = 0$.
And we have the same commutation relations as (\ref{commutationrelationof+-}). \\
              \indent {\it Ghost and matter fields bosonization} ---
The zero mode structure of the ghost $\beta\gamma$ CFT was considered in a different bosonization formula 
about twenty years ago by interchanging the role of $\partial\xi$ and $\eta$ \cite{Horowitz:1988xf}
\begin{eqnarray}
\gamma = - e^{\phi} \partial \xi'=-e^{\phi + \chi'} \partial \chi' \ , ~~ \beta = e^{-\phi} \eta' = e^{-\phi -\chi'} \ .
\label{differentbosonizationformula}
\end{eqnarray}
The authors showed that the superstring scattering amplitudes are manifestly supersymmetric for this bosonization,
in contrast to the original bosonization (\ref{bosonizationformula}).
These two bosonizations, (\ref{bosonizationformula}) and (\ref{differentbosonizationformula}),  
are equivalent for the so called ``small algebras,'' which ignores the zero mode of the fermionic $\xi$ CFT. 
Then the fermionic zero mode of $\eta_0$ does not play any role of consequence. 
There exists one set of zero modes from the scalar field $\phi$. The small algebra of the bosonized fields 
is equivalent to that of the original $\beta\gamma$ CFT. \\ 
        \indent In the ``large algebra,'' which includes the zero mode of the field $\xi$, those two bosonizations, 
(\ref{bosonizationformula}) and (\ref{differentbosonizationformula}), are not equivalent because the $\beta$ and $\gamma$ 
fields have very different roles in the BRST cohomology. In the large algebra of (\ref{bosonizationformula}) 
with $\phi, \xi$ and $\eta$ fields, there is no nontrivial cohomology class 
because the field $\gamma$, which can be written in terms of $\eta$, has a significant role in the BRST cohomology with positive ghost number.
Thus every BRST closed state is also BRST exact state in the bosonization (\ref{bosonizationformula}). 
On the other hand, the bosonization (\ref{differentbosonizationformula}) has a nontrivial cohomology structure.
But it is not unique. Actually there is an infinite number of possible vertex 
operators with different ghost number, which correspond to the same physical state represented by an 
identical cohomology class. \\
          \indent In the context of consistent string theory, there are crucial differences between the matter and ghost 
$\beta\gamma$ CFT bosonizations. The individual matter fields including the matter $\beta\gamma$ CFT do not 
have an independent role in a quantization procedure \cite{Kim:2007pc}, while the ghost 
$\beta\gamma$ CFT has a distinguished role by itself because of the constraint imposed on the ghost number.
Specifically, a similar quantum number of the matter fields has very different role in the sense that the individual 
contribution of the conformal weight of the matter $\beta\gamma$ CFT is always a part of the total matter 
conformal weight. Only the total conformal weight of the matter fields has a constraint and a meaning.   
Another difference is that these two different bosonizations, (\ref{bosonizationformula}) and 
(\ref{differentbosonizationformula}), essentially do not have distinguished features in the matter sector. 
One can choose to have the momentum constraint $p_0 = - k_0$ 
instead of $p_0 = k_0$ which exchange the role between $x^+$ and $x^-$. The BRST operator and the BRST quantization procedure 
are not much different for these two cases because all these zero modes are part of the total matter energy momentum tensor.\\
\indent {\it Application} ---
As it is well known, one immediate application is the bosonization of $\beta\gamma$ CFT into 
the Linear Dilaton theory. 
The basic correspondence between the $\beta\gamma$ CFT and the $\phi\chi$ CFTs is already well known 
in the literature ({\it e.g.} \cite{Polchinski:1998rq}). In this letter we consider the zero modes, in particular, 
and show that there exists a non-compact coordinate after the bosonization. 
The emergence of the Linear Dilaton direction in the bosonized scalar fields $\phi\chi$ CFTs can be 
easily understood from their energy momentum tensor,  
\begin{equation}
T^{\phi\chi} = - \partial \phi \partial \phi/2 + \partial \chi \partial \chi /2 
+ \partial^2 (\chi +(1 - 2 \lambda) \phi) /2   \ . 
\end{equation}
This differs in the last term, which is nothing but the Linear Dilaton direction, from the energy momentum 
tensor of the critical string theory. The fields $\chi$ and $\phi$ always conspire to generate one space-like 
and one time like target manifold coordinates. The same is true for the construction of noncritical string 
theory which can be identified  as Linear Dilaton Theory with the nonlinear sigma model interpretation. \\
\indent These bosonized scalar fields, $\chi = X^0$ and $\phi = X^D$, are part of the full string theory. 
There are other fields, $X^i$ where $ i = 1, \cdots, d$, representing the spatial coordinates. They are 
necessary to cancel the anomaly which comes from the ghost fields. In the action of the theory  
\begin{eqnarray}
S = \frac{1}{2\pi \alpha'} \int d^2 z  \Big(  \partial X^{\mu} \bar{\partial} X^{\nu} \eta_{\mu\nu}
+  \frac{\alpha' R^{(2)}}{2} V_{\mu} X^{\mu} \Big) \ , 
\label{dilatonaction}
\end{eqnarray}
where $R^{(2)}$ is the world sheet scalar curvature.
We can identify the last term as the Dilaton direction $V_{\mu} = (1/2, 0, \cdots , 0, (1-2\lambda)/2 )$, where the 
index $\mu =(0, 1, \cdots ,d, D ) $, covers all the $d+2$ coordinates. 
The vector $V_{\mu}$ satisfies the condition $V_{\mu} V^{\mu} = (24-d)/12$. 
It is not difficult to identify this theory with the known formulation of Linear Dilaton theories 
\cite{Chamseddine:1991qu}. These general theories were already quantized in a light cone gauge \cite{Chamseddine:1991qu}. 
Thus the requirement (\ref{bosonizationcondition}) does not cause any problem in the light-cone gauge quantization. 
The resulting energy dispersion relation for the Linear Dilaton theories is given by \cite{Chamseddine:1991qu}
\begin{equation}
-p^i p_i /2 + 2 p^+ (p^- + 1/2) =  \sum_{m>0} :\alpha_{-m}^i \alpha_m^i: - 1  \ .
\end{equation}
This equation applies generally; even for the case with $p^- = 0$. 
Thus we show that the double bosonized $\beta\gamma$ theory, with appropriate other ingredients, 
are Linear Dilaton theories, which can be quantized in terms of light cone gauge quantization.    \\
              \indent One special case deserves more attention. For $\lambda = 1$, the above action $S$ can be written in the 
light-cone coordinates 
with the identification $X^{\pm} = (X^0 \mp X^D)$. Then the action is similar to the critical string 
theory action with an additional Dilaton term  
$\alpha' R^{(2)} X^+ / 2$. In this case, the symmetries of the action are transparent. Those are  
the translations, rotations and a boost, $SO(8) \times SO(1,1)$. Here $SO(1,1)$ is also 
the symmetry under the Galilean boost transformations, with $X^+$ identified as time 
\begin{eqnarray}
X^i &\longrightarrow& X^i + v^i X^+ \ , \nonumber \\ 
X^- &\longrightarrow& X^- + 2 v^i X_i + v^i v_i X^+ \ .
\end{eqnarray} 
Thus we derive that the bosonized $\beta\gamma$ CFT with additional $d$ 
spatial coordinates are actually the well known Light-like Linear Dilaton theory with 
the light cone coordinate identified as the time coordinate. The quantization is straightforward \cite{Chamseddine:1991qu}. \\
           \indent Let's consider the other side of the story. Before bosonizing, we have the $\beta\gamma$ CFT 
and also additional $d$ scalars for the theory to be consistent. In the conformal gauge, we have the following action. 
\begin{eqnarray}
S' =  \int \frac{d^2z}{2\pi} \left( \beta \bar{\partial} \gamma + \bar{\beta} \partial \bar{\gamma} 
+ \frac{1}{\alpha'}\partial X^i \bar{\partial} X_i  \right) \ . \label{bosonicactionafterbosonization}
\end{eqnarray} 
The commuting matter $\beta\gamma$ CFT has conformal weight $h(\beta) = \lambda$ in addition to the usual anti-commuting 
ghost $bc$ CFT with $h(b) = 2$. $i$ runs from 1 to $d= 26 - 2(6 \lambda^2 - 6 \lambda +1)$ for $X^i$ CFTs. 
This theory is nothing but the ``general Non-Relativistic string theories'' recently proposed \cite{Kim:2007pc}. 
For the general $\beta\gamma$ CFT cases, an explicit quantization 
is not available yet due to the difficulties in constructing vertex operators.
While we don't have many interesting ``supercritical'' bosonic theories with more than critical dimensions , 
there are infinitely many possible supercritical superstring theories \cite{Kim:2007pc}, 
which can be easily be checked with the possible number of spatial coordinates $d' = 8(2-\lambda)$. \\
             \indent Symmetries of the action $S'$ are translations, rotations and boost, $SO(8) \times SO(1,1)$ 
\cite{Kim:2007pc, Gomis:2000bd}. 
$SO(1,1)$ is the symmetry under the Galilean boost transformations   
\begin{eqnarray}
X^i &\longrightarrow& X^i + (v^i / 2) ~ \tau \ ,  \nonumber \\
\beta &\longrightarrow& \beta- v^i~  \partial X^i - (v^i v_i / 4)~  
\partial  \tau  \ , \nonumber \\
\bar{\beta} &\longrightarrow& \bar{\beta} - v^i ~ \bar{\partial} 
X^i - (v^i v_i / 4)~  \bar{\partial} \tau \ , \nonumber
\end{eqnarray}
with a boost parameter $v^i$ and time $\tau = \gamma(z) + \bar{\gamma}(\bar{z}) $. 
Time in the non-relativistic string theory can be generalized to arbitrary linear combinations of $\gamma$ and 
$\bar{\gamma}$ \cite{Kim:2007pc}.   \\
              \indent Thus we establish the basic connection between the Non-Relativistic String theories with the action $S'$
(\ref{bosonicactionafterbosonization}) and the Linear Dilaton theories with action $S$ (\ref{dilatonaction}). 
One pressing task is constructing vertex operators on the Non-Relativistic string theory side. 
For now, with a conservative point of view, we can say the proposed Non-Relativistic String Theories with general 
$\beta\gamma$ CFT \cite{Kim:2007pc} can be quantized and understood in terms of the 
noncritical string theories in arbitrary dimensions \cite{Chamseddine:1991qu} in the same sense that the 
superstring theory could be understood in the bosonized formulation of its ghost system.  \\
            \indent {\it Outlook} --- 
Time-dependent backgrounds in string theory are very interesting because they are intimately related to 
cosmological singularities such as the Big Bang singularity. Non-critical string theories including 
Time-like or Light-like Linear Dilaton theories 
are examples of time-dependent backgrounds in string theory. Despite intensive research, we are far from 
understanding time-dependent backgrounds in string theory because, in some regions of spacetime, 
the string coupling diverges and we lose our ability to analyze the system\cite{Liu:2002ft}. (See some recent developments 
in perturbative approach \cite{McGreevy:2005ci}.) \\  
            \indent We consider the bosonization of the $\beta\gamma$ CFT in the matter sector. Surprisingly, there exists a 
non-compact coordinate in the target manifold, which is consistent with the light-cone quantization. In the full string theory context, 
we establish the basic connection between the proposed non-relativistic string theories \cite{Kim:2007pc} 
and the non-critical string theories \cite{Chamseddine:1991qu}. Due to the non-compact coordinate, it is possible to have 
a full geometric interpretation in non-critical string theories without closed time-like curves. \\
     \indent  There are many remaining tasks to be done on the Non-Relativistic string side. Once we can make progress on the 
Non-Relativistic string theory side, it is clear that this correspondence 
can shed light on many aspects, including the fundamental questions, on the time-dependent string theory backgrounds. 
For example, there seems to be a big discrepancy between the two sides of the correspondence. 
On one side, perturbative string theory breaks down 
in some regions of spacetime because the string coupling becomes large. This makes the general time-dependent backgrounds, 
including Linear Dilaton Theory, difficult to analyze (for example, see {\it e. g.} \cite{Craps:2005wd}). 
On the other hand, we can choose the string 
coupling to be small everywhere in the Non-Relativistic string theories, at least for the known examples \cite{Gomis:2000bd, Kim:2007pc}. 
We view this as a challenge to be resolved in a future work rather than an inconsistency of the bosonization. 
The full potential of this correspondence, between $\beta\gamma$ CFT and non-critical string theories, remains to be seen. \\
   \indent It is pleasure to thank Ori Ganor, Petr Ho\v{r}ava and Ashvin Vishwanath for encouragements, comments and discussions. 
I also thank Anthony Tagliaferro for reading and commenting the final draft. 
This work was supported by Berkeley Center for Theoretical Physics and DOE grant DE-AC02-05CH11231.

\end{document}